\documentclass[12pt]{article}

\usepackage{graphicx}

\begin{document}
\begin{center}
{\Large\bf \boldmath Azimuthal anisotropy and fundamental
symmetries \\ in QCD matter at RHIC}

\vspace*{6mm} {V.A. Okorokov$^a$}\\
{\small \it $^a$ Moscow Engineering Physics Institute (State
University), \\
Kashirskoe Shosse 31, 115409 Moscow, Russian Federation \\
e-mail: VAOkorokov@mephi.ru; Okorokov@bnl.gov}
\end{center}

\vspace*{6mm}

\begin{abstract}
A study of collective behavior in heavy ion collisions provides
one of the most sensitive and promising probes for investigation
of possible formation of new extreme state of strong interacting
matter and elucidating its properties. Systematic of experimental
results for final state azimuthal anisotropy is presented for
heavy ion interactions at RHIC. Experimental data for azimuthal
anisotropy indicate that the final state strongly interacting
matter under extreme conditions behaves as near-ideal liquid
rather, than ideal gas of quarks and gluons. The strong quenching
of jets and the dramatic modification of jet-like azimuthal
correlations, observed in $\mbox{Au+Au}$ collisions, are evidences
of the extreme energy loss of partons traversing matter which
contains a large density of color charges. For the first time,
dependence of the jet suppression on orientation of a jet with
respect to the reaction plane is found at RHIC experimentally. The
model of compound collective flow and corresponding analytic
approach are discussed. The possible violations of $\mathcal{P}$
and $\mathcal{CP}$ symmetries of strong interactions in heavy ion
collisions at different initial energies are considered. Thus, now
the fact is established firmly, that extremely hot and dense
matter created in relativistic heavy ion collisions at RHIC
differs dramatically from everything that was observed and
investigated before.
\end{abstract}

\vspace*{6mm}

Research a heavy ion interactions at high energies and search of a
new state of strongly interacting matter at extremely high density
and temperatures has essential interdisciplinary significance.
Study of collective and correlation characteristics of
interactions allows to obtain new and unique information
concerning various stages of space-time evolution of collision
process, to establish fundamental relation between geometry of
collision and dynamics of final state formation. In heavy ion
collisions, metastable vacuum domains may be formed in the QCD
vacuum in the vicinity of the deconfinement phase transition in
which fundamental symmetries ($\cal{P}$ and/or $\cal{CP}$) are
spontaneously broken. Thus the study of nuclear collisions allows
to investigate one of the most important problem of strong
interaction theory. Relativistic Heavy Ion Collider (RHIC) of
Brookhaven National Laboratory has started to run for physics
2000. RHIC is the world accelerating complex specially designed
and intended entirely for researches in the field of the physics
of strong interactions. The experimental base of RHIC facility
consists of four detectors: small - BRAHMS, PHOBOS and large -
PHENIX, STAR for physics programm with heavy ion beams. Now there
are huge bases of experimental data with high statistics for
$\mbox {Au+Au} (\mbox{Cu+Cu})$ at $\sqrt{s_{NN}} =19.6 (22.4)-200$
GeV, for $\mbox {d+Au}$ and $\mbox {p+p}$ at $\sqrt{s_{NN}} =200$
GeV. Also runs have been executed with small integrated luminosity
for $\mbox {Au+Au}$  at $\sqrt{s_{NN}} =55.8$ GeV, for
$\mbox{p+p}$ at $\sqrt{s_{NN}}=409.8$ GeV, and at low energies:
$\sqrt{s_{NN}}=22 / 9.2$ GeV - for $\mbox{p+p} /
\mbox{Au+Au}$-interactions. Now experimental data are collected on
the large RHIC detectors only.

\section{Azimuthal anisotropy}

One of the most essential features of non-central $\mbox{AA}$
collisions is the violation of azimuthal symmetry of secondary
particle distributions, caused by spatial asymmetry of area of
overlapping of colliding heavy ions. The collective behaviour of
secondary particles manifests itself both in one-particle
$p_{T}$-spectra and in asymmetry of azimuthal particle
distribution with respect to the reaction plane. The first effect
is due to radial azimuthally symmetric expansion, the second
effect is characterized by flow parameters $v_{n}=\langle
\cos\left[n\left(\phi-\Psi_{RP}\right)\right]\rangle$.

The directed flow carries the information about very early stages
of evolution of collision process. The systematic study of the
directed flow $v_{1} $ has been executed in experiments at RHIC at
various $\sqrt{s_{NN}}$ values [1 - 4]. Dependence $v_{1}
\left(\eta(y)-y_{b}\right)$ is presented at Fig.1a at initial
energy range $\sqrt{s_{NN}} =8.8 - 200$ GeV for semi-central heavy
ion collisions, where $y_{b}$ - rapidity of beam particles. The
results obtained at various RHIC energies agree with SPS data [5]
in the region of fragmentation of a beam particle. The most of
transport models underestimates of flow $v_{1}$ for central
rapidity region and agrees with experimental results at the large
values of $|\eta|$ more reasonably. The correlation between the
first and second harmonics indicates on the development of
elliptic flow in plane of event [1]. The recent results show that
the directed flow depends on initial energy but not on the type of
colliding nuclei [4].

It was observed, that the dependences $v_{2}\left(p_{T}\right)$,
integrated $v_{2}$ and differential $v_{2} \left(p_{T}\right)$
parameters on particle mass (type) are described by
phenomenological calculations on the basis of hydrodynamics up to
$p_{T} \sim 2$ GeV/$c$ well enough for different particles [6, 7].
The indicated above range of $p_{T}$ contains $\sim 99\%$ of
secondary particles. Thus, the global dynamic feature of the
created matter is the collective behaviour described in the
framework of relativistic hydrodynamical model at qualitative
level. The increasing of $v_{2}$ and systematic decreasing of
$v_{2} \left(p_{T}\right)$ with growth of particle masses is the
additional and essential indication on presence of the common
velocity fields. Dependence $v_{2} \left(\sqrt{s_{NN}}\right)$ is
presented on the Fig.1b based on [8, 9]. One can see the $v_{2}
\left(\sqrt{s_{NN}}\right)$ increases smoothly at
$\sqrt{s_{NN}}>5$ GeV. The flow energy dependence was fitted by
function
$a_{0}+a_{1}\left[\sqrt{s_{0}}/\left(\sqrt{s_{NN}}-2m_{p}\right)\right]
^{a_{2}}+a_{3}\lambda^{a_{4}}+a_{5}\lambda^{a_{6}}+a_{7}\left[\ln\lambda
\right]^{a_{8}}$ ($\lambda \equiv s_{NN}/s_{0},~s_{0}=1
\mbox{GeV}^{2}$, $m_{p}$ -- proton mass) which is similar to that
for $\sigma_{tot}^{pp}$ approximation [10]. This function agrees
with all available data reasonably, but the approximation with
best statistical quality $\left(\chi^{2}/ndf=3.83\right)$ shows
the decreasing in TeV energy range (Fig.1b, curve 1). There is
some poor quality $\left(\chi^{2}/ndf=4.83\right)$ for curve 2
(Fig.1b) which shows a reasonable behaviour for all energy domain
understudy. One can see the experimental data for both ultra-high
energy range (LHC) and for intermediate energies
$\sqrt{s_{NN}}=5-50$ GeV (FAIR, NICA) are essential for more
unambiguous approximation of $v_{2} \left(\sqrt{s_{NN}}\right)$.
As seen, the PHOBOS point at $\sqrt{s_{NN}}=19.6$ GeV agrees well
with the results observed at close SPS energies. It is important
to note, that at high RHIC energies the hydrodynamical limit for
elliptic collective flow parameter is reached for the first time.
The dependence of $v_{2}$ on centrality emphasizes additionally of
importance of more exact knowledge for initial state in a
nuclei-nuclear collisions for the correct description of final
state matter evolution. Essential nonzero value and the basic
dependences for $v_{2}$ are observed for various types of
colliding (symmetric) heavy ion beams [11, 12]. Dependence
$v_{2}(\eta)$ has been obtained at RHIC for various initial
energies both for $\mbox{Au+Au}$, and for $\mbox{Cu+Cu}$
collisions.
\begin{figure}[h]
\begin{center}
\begin{tabular}{cc}
\mbox{\includegraphics[width=7.5cm,height=7.0cm]{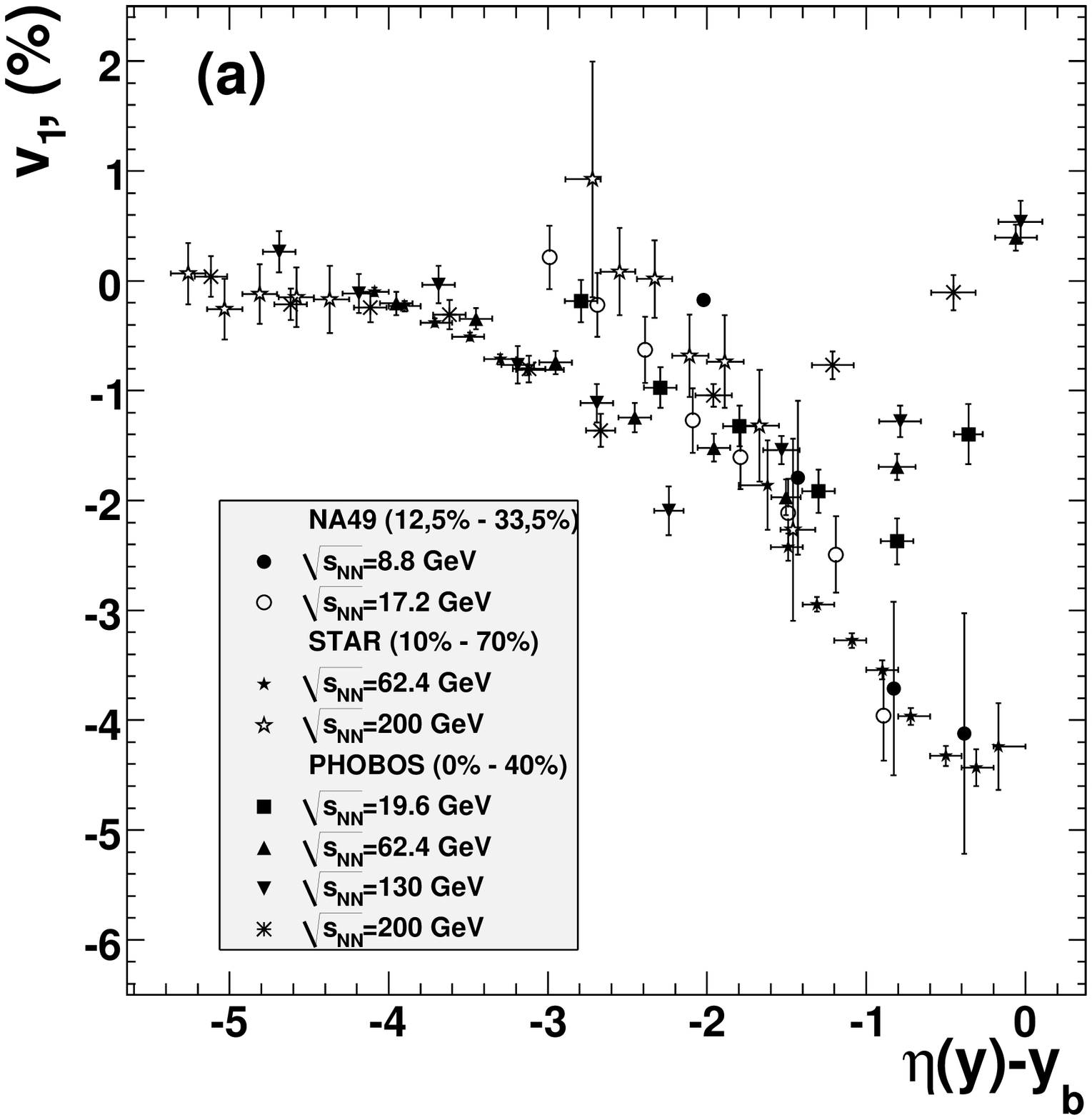}}&
\mbox{\includegraphics[width=7.5cm,height=7.0cm]{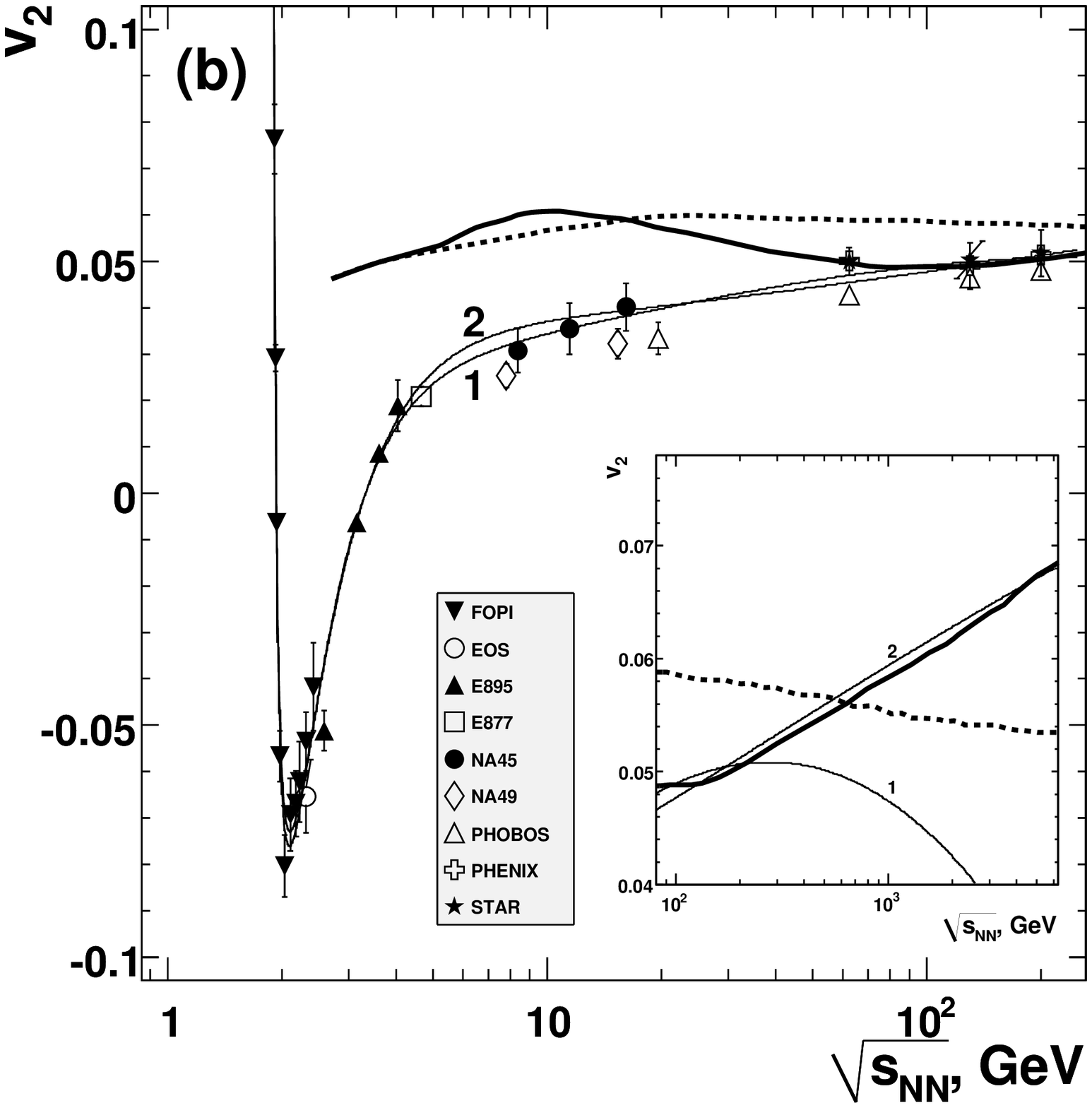}}
\end{tabular}
\end{center}
\caption{(a) Parameter $v_{1}$ for the charged particles at SPS
($\mbox{Pb+Pb}$) and RHIC ($\mbox{Au+Au}$) energies. Collision
centralities are specified. (b) Energy dependence of $v_{2}$
parameter for midrapidity domain in heavy ion collisions. A thick
solid curve corresponds to the hydrodynamical calculations for
strongly interacting matter EOS with phase transition, a dashed
curve -- for the EOS of hadron gas without phase transition. Inner
picture shows predictions up to the ultra-high energies (see text
for more detail). Statistical errors are indicated only.}
\end{figure}
The different phenomenological models describe the $v_{2}(\eta)$
at qualitative level. Essential excess for partonic cross-sections
used in such models above pQCD predictions can indicates on
significant non-perturbative effects in the partonic matter formed
on the RHIC.

Experimental $v_{2}$ results, obtained both for charged hadrons
and for the identified particles of various types, allow to make a
choice in favour of the equation-of-state (EOS) of strongly
interacting matter with presence a quark-gluon phase at early
stages of space-time evolution of the formed medium with the
subsequent transition in hadronic phase (Fig.1b). Moreover the
hybrid model with jet quenching, unlike hydrodynamics, predicts
saturation of $v_{2}$ in the intermediate $p_{T}$ domain with the
subsequent decreasing at higher transverse momentum and describes
dependence $v_{2}\left(p_{T}\right)$ for $p_{T}>2$ GeV/$c$ at
large density of gluons $dN{g}/dy \sim 10^{3}$ qualitatively [6].
This experimental result (together with other ones) is the direct
evidence of hot and dense matter formation in which there are
partonic hard scattering and finite energy loses of partons and
products of their fragmentation at traversing of this medium. The
comparative analysis of $\mbox{p+p},~\mbox{d+Au}$ and $\mbox
{Au+Au}$ collisions at energy $\sqrt{s_{NN}}=200 $ GeV has
demonstrated that there is a significant contribution from
anisotropic (elliptic) flow namely in non-central nuclei-nuclear
collisions up to, at least, $p_{T} \simeq 10$ GeV/$c$ [13]. The
large elliptic flow signal indicates on the one hand on the fast
achieving of thermodynamic equilibrium at RHIC energies, and on
the other hand - the medium contituents should interact with each
other intensively enough at the early stages of space-time
evolution of matter already, that more corresponds to conditions
of a liquid, than gas.

At RHIC experimental results have been obtained for parameter of a
collective elliptic flow for identified $\pi^{0}$-mesons and
inclusive $\gamma$ in $\mbox{Au+Au}$ collisions at
$\sqrt{s_{NN}}=200$ GeV. The obtained experimental indication on
small $v_{2}$ values for direct photons makes preferable the naive
scenario of direct photon production in processes of the hard
scattering, which occurs at the earliest stages of space-time
evolution of the created matter [14]. The universal dependence of
parameter $v_{2}/n_{q}$ on normalized kinematic variables
($p_{T}/n_{q}, E_{kT}/n_{q}, n_{q}$ -- constituent quark number)
is observed for a wide set of secondary mesons and baryons [12,
15]. This scale behaviour is the experimental evidence of presence
of an essential collective partonic flow and similar character of
$s$-quark flow with elliptic flow of light $u, ~d$-quarks.

The experimental RHIC data for elliptic flow $v_{2}$ for light
flavour particles assume that final state matter is characterized
by small free path length of constituents in comparison with the
sizes of system and value of $\eta_{s}/s \sim 0.1$ ($\eta_{s}$ -
shear viscosity, $s$ - entropy density) is close to the bottom
quantum limit for strong coupling systems in the energy domain
$\sqrt{s_{NN}} \sim 100$ GeV. The elliptic anisotropy of the
particles with heavy flavour quarks has been investigated in
$\mbox{Au+Au}$ collisions at $\sqrt{s_{NN}}=200$ GeV. This result
together with results for nuclear modification parameter
$R_{AA}\left(p_{T}\right)$ is strong evidence in favour of the
conclusion about strong coupling of heavy quarks with the final
state matter [16]. Experimental values of $v_{2}^{HF}$ are larger
essentially of model predictions based on pQCD. The Langevin
transport model with small relaxation times and / or small
diffusion coefficients of heavy quarks $K_{d}^{HQ}$ allows to
obtain a reasonable agreement at a qualitative level between
experimental and calculated flow values. Estimations for
$\eta_{s}/s$, obtained in such way, are close to the bottom
quantum limit also and these estimations agree with results for
sector of light quarks. Thus, interpretation of experimental RHIC
results for azimuthal anisotropy on the basis of hypothesis about
creation at early stage (quasi)ideal partonic liquid is the most
proved at present. The alternative approach to explanation of
small $\eta_{s}/s $ value is based on the  presence of anomaly
shear viscosity $\eta_ {s}^{A}$, arising due to turbulence of the
color magnetic and electric fields generated by expanding
quark-gluon system [17].

The higher order even harmonics have been obtained in STAR
experiment for charged particles in $\mbox{Au+Au}$ collisions at
initial energy $\sqrt{s_{NN}}=200$ GeV. Values of even harmonics
for charged particles, averaged over $p_{T}$ and $\eta$
$\left(|\eta|<1.2\right)$ for minimum bias events, are equal
$(\%)$: $v_{2} =5.180 \pm 0.005;~v_{4} =0.440 \pm
0.099;~v_{6}=0.043 \pm 0.037;~v_{8} =-0.06 \pm 0.14 $ [18].

Identification of jets in relativistic heavy ion interactions was
carried out on a statistical basis so far. Two peaks are observed
in experimental correlation functions $C_{2}(\Delta \phi) \propto
\int d\Delta \eta N\left(\Delta\phi,\Delta\eta\right)$ for
$\mbox{AA}$ collisions, which correspond two-jet event structure.
The significant suppression of peak at large relative azimuthal
angles $(\Delta \phi \simeq \pi)$ was observed in azimuthal
correlations of two particles with high $p_{T}$ for central
$\mbox{Au+Au}$ in comparison with $\mbox{p+p}$. Moreover, this
effect increases with centrality increasing in nuclei-nuclear
interactions. The experimental observations have been interpreted
as the critical evidences of large losses of parton energy in
dense deconfinement matter, predicted by pQCD as a state of
quark-gluon plasma. Control experiments with $\mbox{d+Au}$
collisions have shown that the back-to-back peak is present in
this case and its characteristics are close to ones which are
observed in $\mbox{p+p}$ interactions. Therefore, experimental
RHIC results on two-hadron azimuthal correlations are one of the
most obvious and important evidences in favour of creation of new
state of strongly interacting matter at final stage of central
nuclei-nuclear collisions. This final state matter is
characterized by large energy losses and by opacity for partons
with high $p_{T}$ and for products of their fragmentation. The
subsequent study at higher $p_{T}$ both for trigger particles and
for associated ones have allowed to find clear two peaks in
$C_{2}(\Delta \phi)$, corresponding two-jet event structure both
in semicentral and even central $\mbox{Au+Au}$ collisions as well
as in $\mbox{d+Au}$. As expected, the back-to-back peak is
suppressed essentially in central $\mbox{Au+Au}$ events (as well
as at smaller $p_{T}$) [19]. This is the first direct experimental
observation of two-jet event structure in central $\mbox{AA}$
collisions, corresponding to pQCD predictions at qualitative
level. Researches of back-to-back peak characteristics have been
executed for collisions with various ion types and initial
energies [20]. These experimental results indicate on the
essential response of medium on energy deposition of hard parton
traversed it. At present there are a several scenarios for medium
response on hard jet traversing. It seems the experimental results
agree with shock wave model at qualitative level only [20, 21].
Thus, additional experimental data and theoretical study are
necessary for more unambiguous observation and explanation of cone
topology in two dimensions.

Fig.2 shows experimental results for investigations of hadron jet
correlations with respect to the reaction plane at SPS energy [22]
and RHIC one [13] in $\mbox{Pb+Au}$ and $\mbox{Au+Au}$ collisions,
accordingly.
\begin{figure}[h]
\centerline{\includegraphics[width=9.0cm,height=6.6cm]{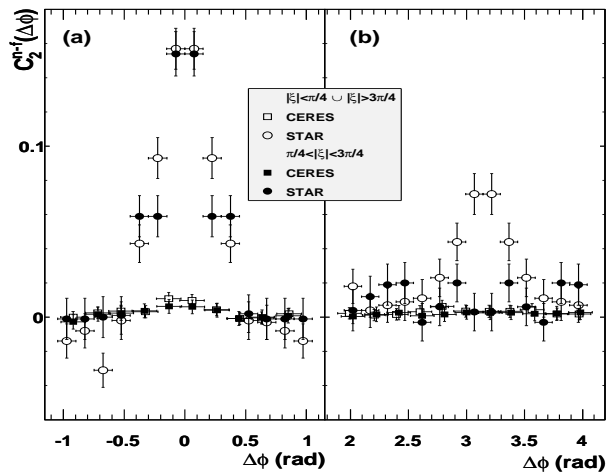}}
\caption{Azimuthal two-particle correlation functions at SPS and
RHIC energies in the ranges of small (a) and large (b) relative
azimuthal angles. The open symbols correspond to emission of
trigger particles in-plane, solid symbols -- out of plane
direction for $\mbox{AA}$ interactions. Statistical errors are
presented only.}
\end{figure}
The jet was considered as directed in reaction plane, if
$|\xi|<\pi/4 \bigcup |\xi|>3\pi/4$, and directed out of reaction
plane, if $\pi/4<|\xi|<3\pi/4$, where $\xi \equiv
\phi^{tr}-\Psi_{2} $ -- relative azimuthal angle between a
reaction plane estimated by angle of second order event plane
$(\Psi_{2})$, and a trigger particle. Experimental data obtained
at various initial energies, for small and large relative
azimuthal angles domain are compared on Fig.2a, 2b, accordingly.
As seen, excess over a level of elliptic flow is much weaker at
SPS energy $\sqrt{s_{NN}} =17.3$ GeV, than for RHIC energy in the
$\Delta \phi \sim 0$ domain (Fig.2a). At $\Delta \phi \sim \pi$
the values of SPS correlation functions are close to zero level
(Fig.2b). One can see correlation functions for various directions
of a trigger particle agree with each other well at SPS energy for
both cases $\Delta \phi \sim 0$ (Fig.2a) and for $\Delta \phi \sim
\pi$ (Fig.2b), that essentially differs from behaviour of jet-like
correlations on RHIC in case of hadron jets passed inside medium
with various directions with respect to the reaction plane.
Therefore, suppression effect at RHIC is stronger significantly
for hadron jet traversing of hot and dense matter on direction out
of reaction plane, and suppression is weaker in the case of jet
passing inside a matter on direction in-plane. This RHIC result is
the first experimental evidence of strong correlation of
suppression strength of hard hadron jets traversing the volume of
hot and dense strongly interacting matter, with path lengths which
are passed by jets inside of this matter.

The model of (multi)component collective elliptic flow has been
suggested in [23 - 25] based on the experimental RHIC data for
azimuthal anisotropy. In the framework of this model the
correlations with respect to the reaction plane are supposed both
for soft particles, and for hard particles (hadronic jets). The
model of compound flow takes into account the number of jets per
event, average multiplicity per jet, dependence of jet yield on
the orientation with respect to the reaction plane, and
independent "soft" particle production. The generalized formulas
were derived for two-particle distribution on a relative azimuthal
angle and for two-particle distributions in / out with respect to
the reaction plane [23 - 25]. These analytic calculations provide
the framework for a consistent description of the elliptic flow
measured via the single-particle distribution with respect to the
reaction plane, jet yield per event, and the amplitude of
flow-like modulation in the two-particle distribution in the
relative azimuthal angle. It seems, the difference between soft
particle flow and parameter of jet correlations with respect to
the reaction plane is (very) close to zero at RHIC energies. But
jet production will give a more significant contribution at higher
(LHC) energies and this difference may be more visible. The model
of compound flow agree with expectations, that energy losses of
partons are sensitive to energy-momentum tensor $T^{\mu \nu} =
(e+p)u^{\mu}u^{\nu}-pg^{\mu \nu}$, described the global properties
of created matter. Hence, partons and products of their
fragmentation can be sensitive not only to the equation-of-state,
but to the common velocity fields of collective flow $u^{\mu}$
also, i.e. hard jets appear "enclosed" in collective expansion of
surrounding medium at all.

\section{Fundamental symmetries in QCD-matter}

The possible $\mathcal{P}$ and/or $\mathcal{CP}$ violation in
strong interactions requires the deconfinement state of matter
with restored chiral symmetry [26, 27]. Thus the positive and
reliable experimental results for $\mathcal{P}$ and/or
$\mathcal{CP}$ violation in the strong interactions would be prove
the clear evidence of deconfinement and chirally symmetric phase
creation and establish experimentally the presence of topological
configurations of gluon fields and their role in chiral symmetry
breaking [27]. The lattice calculations show the non-trivial
topological structure of gluon fields which can be characterized
by topological charge (winding number) $Q_{W}=\eta\int
d^{4}xG^{a}_{\mu\nu}\tilde{G}^{\mu\nu}_{a},~\eta \equiv
g^{2}_{s}/32\pi^{2}$. The gauge (color) fields with non-zero
$Q_{W}$ induce difference between number of left-and right-handed
fermions. The topological charge changing transitions are
exponentially suppressed at zero temperature (instanton modes).
But the such transitions unsuppressed at finite temperature
(sphaleron modes). In chiral limit the charge difference will be
created between two sides of a plane perpendicular to the
chromo-magnetic field. Thus one can define chiral magnetic effect
as a charge separation by gauge field configurations with non-zero
$Q_{W}$ in the presence of background (electric) magnetic field.
The clear equation for effect strength was derived in [28]. One
needs to emphasize two features which are essential for
experimental investigations namely. First of all chiral magnetic
effect corresponds the early collision stages because of magnetic
field falls off rapidly. Secondly, electrical charge is conserved
in hadronization. Thus charge separation for quarks implies charge
separation with respect to the reaction plane for hadron states
observed experimentally.

The presence of non-zero angular momentum in non-central
$\mbox{AA}$ collisions is equal of the external magnetic field.
Thus the experimental investigation of electrical charge
separation of produced particles in non-central heavy ion
collisions makes it possible to study $\mathcal{P}$ and/or
$\mathcal{CP}$-odd domains. A correlator, directly sensitive to
$\mathcal{P}$-event quantitative parameter is $\langle
\cos\left(\phi_{\alpha}+\phi_{\beta}-2\Psi_{RP}\right)\rangle$
[29]. The three-particle correlations allow to use a following
characteristic: $\langle
\cos\left(\phi_{\alpha}+\phi_{\beta}-2\phi_{\gamma}\right)\rangle=\langle
\cos\left(\phi_{\alpha}+\phi_{\beta}-2\Psi_{RP}\right)\rangle
v_{2}^{\gamma}$, which is more useful for experimental
applications [18]. The preliminary experimental results were
obtained by STAR experiment at RHIC for $\mbox{Cu+Cu}$ and for
$\mbox{Au+Au}$ collisions at initial energies 62.4 and 200 GeV
[30]. The signal of charge separation in $\mbox{Cu+Cu}$ collisions
is some larger than that for $\mbox{Au+Au}$ for the same
centralities and the experimental signal has a typical hadronic
"width" [30]. This feature agrees qualitatively with scenario of
stronger suppression of the back-to-back correlations in heavier
$(\mbox{Au+Au})$ collisions. Thus the preliminary STAR
experimental results agree at qualitative level with the magnitude
and main features of the theoretical predictions for $\mathcal{P}$
–violation in nuclear collisions at RHIC [30]. At present there is
no indication on the some other effects which could be imitate the
experimental signal and the fundamental symmetry violation in
heavy ion collisions. But one need more rigorous study of such
possible imitation effects for more unambiguous conclusion. The
energy dependence of strength of parity violation effect is not
trivial. It seems the necessary conditions can be reached in
nuclei-nuclear collisions in sufficiently wide energy range. The
some important medium properties are very sensitive to the
vicinity of phase transition boundary. Therefore, perhaps, the
effect of hypothetic fundamental symmetry violation in strong
interaction might be more clear at lower energies [27].

\section{Summary}

The main RHIC results and corresponding phenomenological models
are presented for two-particle azimuthal correlations.
Investigation of azimuthal anisotropy has allowed to obtain a set
of the important experimental results for the first time. Fast
space-time evolution is observed for volume of the final state
matter. This evolution is described by significant gradients of
pressure at early stages. Azimuthal anisotropy agrees well enough
with calculations in the framework of hydrodynamical models with
phase transition and very small viscosity for the small transverse
momentum domain. The manifestation of scale properties predicted
by model of quark coalescence, significant values of elliptic flow
for multistrange baryons and heavy flavour quarks allow to assume,
that the collective behaviour on partonic stages is observed. The
final state matter is opaque for partons and products of their
fragmentation passed inside of it, and the degree of suppression
depends on thickness of a layer of a traversed matter
significantly. The generalized formulas have been derived for
azimuthal correlation functions with taking into account various
components (soft and hard) of elliptic collective flow and for
different (in/out) orientations with respect to the reaction
plane. These equations are based on experimental RHIC results for
azimuthal asymmetry for both soft and hard particles. Thus, the
matter created in a final state of nuclei-nuclear collisions at
high RHIC energies, differs qualitatively from all matter state
created and investigated in laboratory conditions early, and color
degrees of freedom namely are adequate ones for the description of
the early stages of space-time evolution of the RHIC matter. The
created matter is rather similar on (quasi)ideal liquid of color
constituents (partons), than on (quasi)ideal quark-gluon gas, as
expected earlier. The preliminary experimental results indicate
the possible $\cal{P}$-violations in nuclear collisions at RHIC.
This observation qualitatively agree with theoretical expectations
for fundamental symmetry violations in QCD-matter under extreme
conditions. At present the new direction of investigations is
formed intensively which can be designated as "relativistic
nuclear physics of condensed / continuous matter".


\end{document}